\documentclass[twocolumn]{aastex62}

\usepackage{amsmath}
\graphicspath{{./}{figures/}}
\usepackage{natbib}

\submitjournal{ApJ Letters}

\shorttitle{The size-luminosity relation in the Coma cluster}
\shortauthors{Danieli \& van Dokkum}

\begin{document}

\newcommand\XXX[1]{{\textcolor{red}{\textbf{XX\ #1\ XX}}}}

\title{Revisiting the Size-Luminosity Relation in the Era of Ultra Diffuse Galaxies}

\correspondingauthor{Shany Danieli}
\email{shany.danieli@yale.edu, shanyi1@gmail.com}

\author[0000-0002-1841-2252]{Shany Danieli}
\affil{Department of Physics, Yale University, New Haven, CT 06520, USA \\}
\affil{Yale Center for Astronomy and Astrophysics, Yale University, New Haven, CT 06511, USA \\}
\affil{Department of Astronomy, Yale University, New Haven, CT 06511, USA \\}

\author[0000-0002-8282-9888]{Pieter van Dokkum}
\affiliation{Department of Astronomy, Yale University, New Haven, CT 06511, USA \\}

\begin{abstract}
Galaxies are generally found to follow a relation between their size and luminosity, such that luminous galaxies typically have large sizes.  The recent identification of a significant population of galaxies with large sizes but low luminosities (``ultra diffuse galaxies", or UDGs) raises the question whether the inverse is also true, that is, whether large galaxies typically have high luminosities.  Here we address this question by studying a {\em size}-limited sample of galaxies in the Coma cluster. We select red cluster galaxies with sizes $r_{\mathrm{eff}} > 2 \ \mathrm{kpc}$ down to $M_{g} \sim -13 \ \mathrm{mag}$ in an area of $9 \ \mathrm{deg}^2$, using carefully-filtered CFHT images. The sample is complete to a central surface brightness of $\mu_{g,0}\approx 25.0 \ \mathrm{mag\,arcsec}^{-2}$ and includes 90\,\% of Dragonfly-discovered UDGs brighter than
this limit. Unexpectedly, we find that red, large galaxies have a fairly uniform distribution in the size-luminosity plane: there is no peak at the absolute magnitude implied by the canonical size-luminosity relation. The number of galaxies within $\pm 0.5$ magnitudes of the canonical peak ($M_g = -19.69$ for $2<r_{\mathrm{eff}}<3$\,kpc) is a factor of $\sim 9$ smaller than the number of fainter galaxies with $-19<M_g<-13$. Large, faint galaxies such as UDGs are far more common than large galaxies that are on the size-luminosity relation. An implication is that, for large galaxies, size is not an indicator of halo mass. Finally, we show that the {\em structure} of faint large galaxies is different from that of bright large galaxies: at fixed large size, the S\'ersic index decreases with magnitude following the relation $\log_{10} n \approx -0.067M_g-0.989$.


\end{abstract}



\section{Introduction} \label{sec:intro}

\begin{figure*}[t!]
{\centering
  \includegraphics[width=180mm]{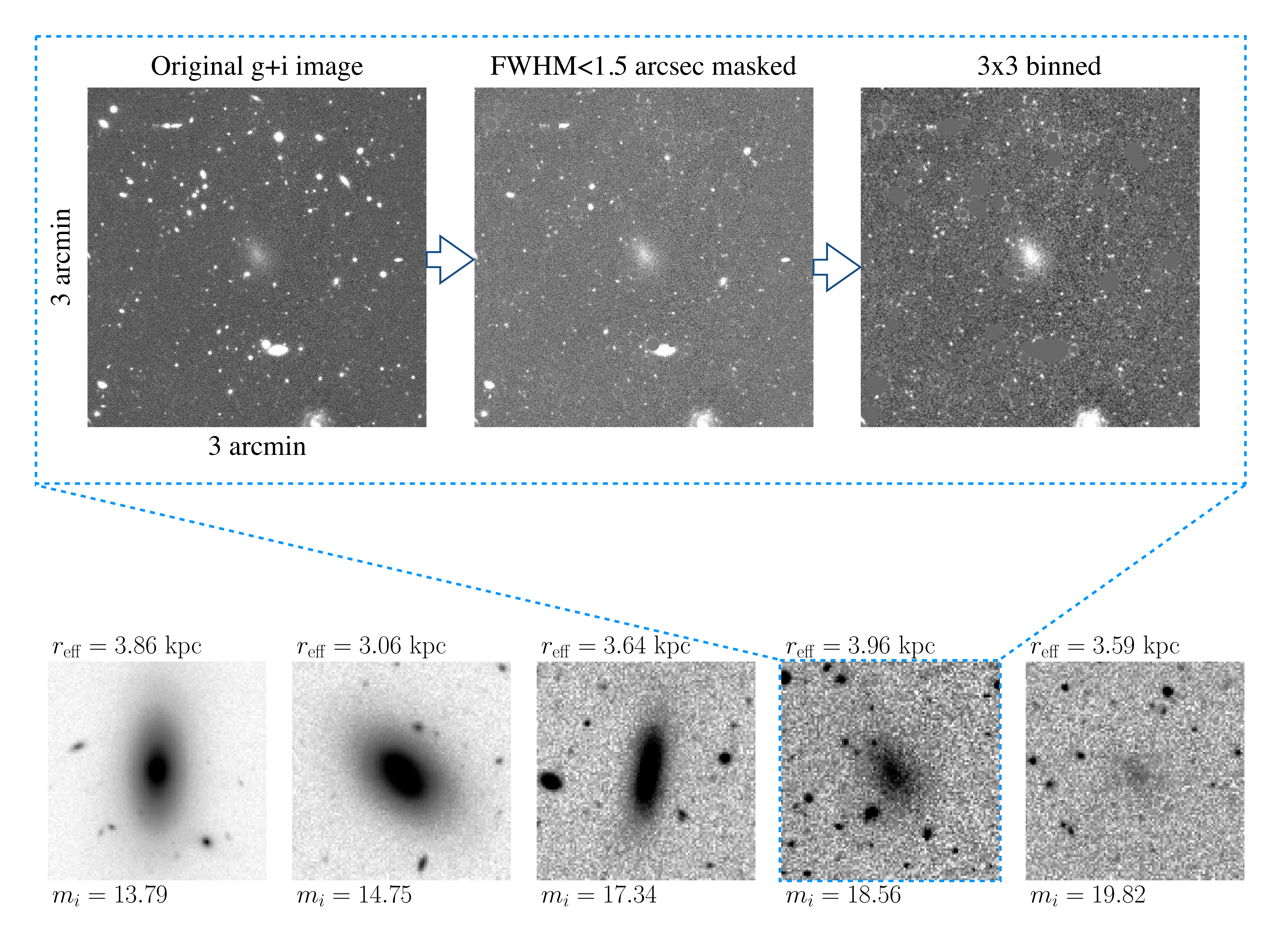}
  \caption{Co-added $g+i$ CFHT images, demonstrating the steps of the detection pipeline used for creating the catalog (upper panel). The panels show a $3'\times{}3'$ area centered on the UDG Dragonfly~44 (van Dokkum et al.\ 2015b), one of the large low surface brightness galaxies in our sample. Compact objects are masked and the image is binned to facilitate the detection of low surface brightness emission. In the lower panel we show examples for galaxies detected with our detection pipeline. All have similar effective radii of $3-4 \ \mathrm{kpc}$ but they span a factor of $\sim 250$ in luminosity, ranging from 13.8 to 19.8 mag.}
  \label{fig:data}
}
\end{figure*}


Early-type and late-type galaxies exhibit scaling relations, involving their structural, photometric and physical parameters (\citealt{1976ApJ...204..668F}, \citealt{1977A&A....54..661T}, \citealt{1987ApJ...313...59D}). These relations are used as distance indicators (e.g. \citealt{2001ApJ...553...47F}), provide insights into the formation and evolution of galaxies (e.g. \citealt{2007MNRAS.374.1479G}), as well as constraints on the nature of dark matter (e.g. \citealt{2002ARA&A..40..263S}).
One of the most straightforward relations is that between the sizes of galaxies (parameterized by the half-light or half-mass radius, $r_{\mathrm{eff}}$) and their luminosities or masses (\citealt{1977ApJ...218..333K}). The slope and normalization of this relation presents distinct trends depending on environment and cosmic time.
Previous studies have derived the size-luminosity relation from large datasets complete down to faint limits, carefully measuring the photometric parameters of galaxies using advanced galaxy modeling and fitting (\citealt{2003MNRAS.343..978S}, \citealt{2014MNRAS.443..874B}). 
The fairly tight log-linear relation between the size and the integrated magnitude of galaxies that is found in these studies suggests that the majority of large galaxies are bright.

Recent low surface brightness imaging efforts resulted in the discovery of a significant population of galaxies with large effective radii ($r_{\mathrm{eff}} \gtrsim 1.5 \ \mathrm{kpc}$) and low central surface brightness ($\mu_{0} \gtrsim 24 \ \mathrm{mag \ arcsec}^{-2}$), mostly in cluster environments (e.g. \citealt{2015ApJ...798L..45V}, \citealt{2015ApJ...809L..21M}, \citealt{2015ApJ...807L...2K}).
Some isolated examples of these large, low surface brightness galaxies, dubbed ``ultra diffuse galaxies" (UDGs), were detected earlier (e.g. \citealt{1997AJ....114..635D}, \citealt{2006ApJ...651..822C}) but it is their large abundances, particularly in cluster environments, that is new to us.
In light of this discovery it is interesting to examine how UDGs modify the derived galaxy size-luminosity relation. Speficially, the key questions are whether most large galaxies are, in fact, luminous, and whether there is a continuum in luminosity between large ``normal'' galaxies and UDGs.

Here we present the size-luminosity relation of galaxies in the Coma Cluster, using imaging from the Canada France Hawaii Telescope (CFHT) covering an area of $9\ \mathrm{deg}^2$, down to low surface brightness levels. 
To ensure size completeness, we focus on galaxies with large effective radii ($r_\mathrm{eff} > 2 \ \mathrm{kpc}$).
We show that  large galaxies do not follow the canonical size-luminosity relation and that the apparently tight size-luminosity relation might be a result of selection effects, due to poor sensitivity to large, low surface brightness galaxies. 
Throughout this Letter, we assume a distance of $100 \ \mathrm{Mpc}$ to the Coma cluster and a flat $\Lambda \mathrm{CDM}$ model with parameters $\Omega_{\mathrm{m}} = 1-\Omega_{\Lambda} = 0.27$, $\Omega_\mathrm{b} = 0.0469$, $h=H_0/(100 \ \mathrm{km} \ \mathrm{s}^{-1} \ \mathrm{Mpc}^{-1}) = 0.7$, $\sigma_8 = 0.82$, and $n_\mathrm{s} = 0.95$ compatible with combined constraints from the \textit{Wilkinson Microwave Anisotropy Probe}, Baryonic Acoustic Oscillations, supernovae, and cluster abundance (\citealt{2011ApJS..192...18K}).


\section{Analysis} \label{sec:analysis}

\begin{figure*}[t!]
{\centering
  \includegraphics[width=150mm]{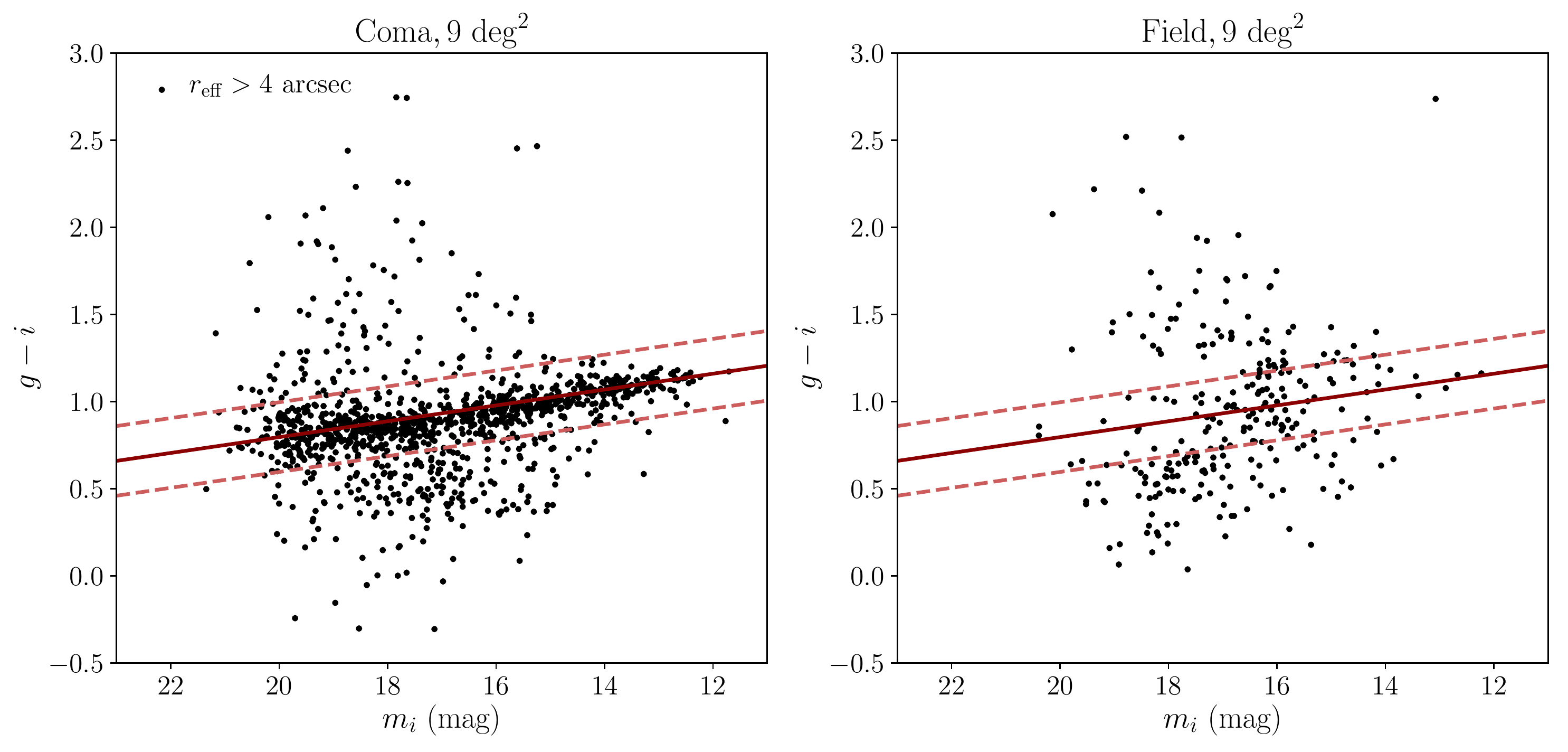}
  \caption{Color-magnitude diagram for our sample of galaxies with $r_{\mathrm{eff}}>4\arcsec$ in the Coma cluster field (left) and in the random field (right), both covering an area of $9 \ \mathrm{deg}^2$. The solid red line and dashed pink lines shows our fit to the red sequence and limits of 0.2 magnitudes above and below the fitted red sequence, respectively. In the following, only galaxies within these limits are considered in the analysis.}
  \label{fig:CMD}
}
\end{figure*}

\begin{figure*}[t!]
{\centering
  \includegraphics[width=150mm]{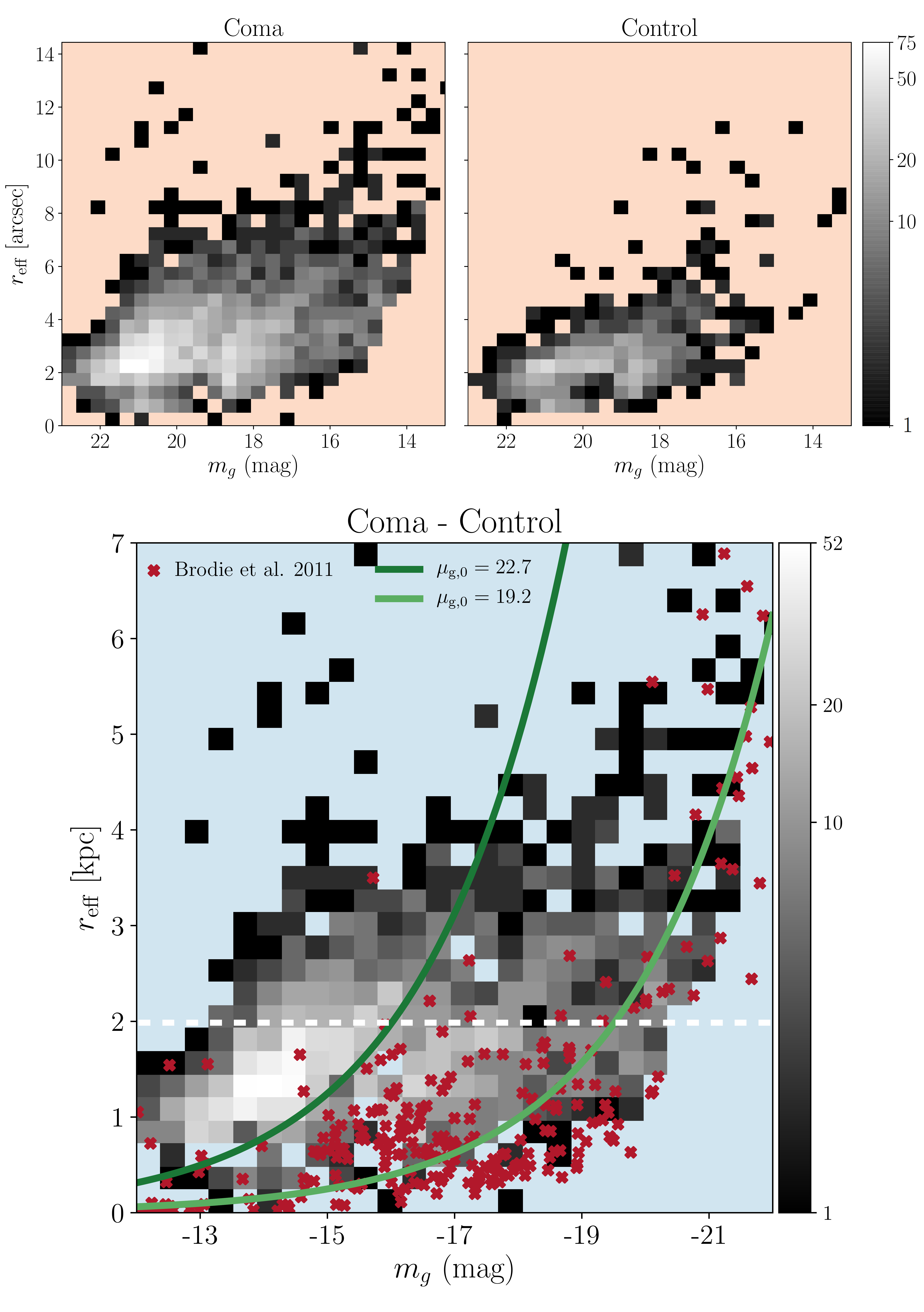}
  \caption{The distribution of red galaxies in the size-magnitude plane, for galaxies with $r_{\mathrm{eff}} > 2 \ \mathrm{kpc}$. The Coma field is shown at top left and the control field at top right.  The greyscale indicates the number of galaxies at each location. In the bottom panel the field counts are subtracted from the Coma counts, correcting for field contamination. The corrections are small. The dashed white line marks the size completeness of the sample. 
Red points denote a class of quiescent objects from Brodie et al. (\citeyear{2011AJ....142..199B}). Large galaxies in this sample have an approximately constant central surface brightness of $19.2 \ \mathrm{mag} \ \mathrm{arcsec}^{-2}$ (light green line). The population of faint and large galaxies with central surface brightness fainter than $22.7 \ \mathrm{mag} \ \mathrm{arcsec}^{-2}$ (dark green line) is missing in the Brodie et al.\ compilation.
The magnitude distribution of large galaxies is much broader than was suggested by the Brodie sample; this is the central result of this paper.}
  \label{fig:scatter}
  }
\end{figure*}

\subsection{Data and Catalog Selection} \label{sec:data}
Our size-limited sample of Coma galaxies is constructed from imaging data obtained with the
Canada France Hawaii Telescope (CFHT), as they cover a large region of the Coma cluster in a homogeneous way and it was shown in van Dokkum et al. (\citeyear{2015ApJ...798L..45V}) they reach the require depth.
We obtained the data from the Canadian Astronomy Data Centre. A $3^{\circ} \times 3^{\circ}$ field was imaged with a 9-pointing mosaic, in the $g$ and $i$ bands (\citealt{2014MNRAS.440.1690H}). Exposure times were short, at 300 s per pointing per filter. The image quality is FWHM $\approx 0\farcs 8$ and the sampling is $0\farcs 186 \ \mathrm{ pixel}^{-1}$.
We also obtained images in a $3^{\circ} \times 3^{\circ}$ area in the Canada-France-Hawaii-Telescope Legacy Survey Wide fields (W1) (CFHTLS; \citealt{2013MNRAS.433.2545E}). These data are used to correct the Coma counts for contamination, as detailed below. Similarly to the Coma data, 9-pointing mosaic images were obtained in the $g$ and $i$ bands. 
Exposure times were much longer, at 3000\,s and 4300\,s per pointing for the $g$ and $i$ filters respectively. For each of the 9 images we measured the noise, using the distribution of flux values measured in $25\times 25$ pixel boxes placed in empty areas. For each Coma field we assigned a control field whose variation in $25\times 25$ pixel boxes was matched to that in Coma. This matching was done by adding Poisson-distributed noise to the control field.

To create a size-limited sample, we need to include both luminous galaxies and very faint low surface brightness objects. We created this ``high dynamic range" catalog in the following way. The same procedure was applied to both the Coma images and the blank field data.
First, the $g$ and $i$ images were summed to increase the S/N ratio. Then,
SExtractor (\citealt{1996A&AS..117..393B}) was used to create a catalog of bright objects, by
setting the detection threshold to $>5 \sigma$ per pixel. Objects with high measured flux ($\mathrm{FLUX\_AUTO} > 10^4$, corresponding to $gi<20.7$ mag) were included in the catalog and then masked in the images, in preparation for the next steps. Objects with FWHM smaller than $1.5 \ \mathrm{arcsec}$ were not included in the catalog but masked outright. Next,
the masked image was rebinned by $3 \times 3$ to a lower resolution to increase the S/N ratio per pixel for a better detection of low surface brightness objects. 
SExtractor was run a second time on this masked and re-binned image, now using a lower detection threshold of  $3 \sigma$ per pixel. Objects with FWHM smaller than $3.5 \ \mathrm{arcsec}$ and objects that are likely to be stars ($\mathrm{CLASS\_STAR} > 0.9$) were masked. 
SExtractor was run a third time after this masking step, with a low detection threshold of $2 \sigma$ per pixel and a large minimal number of detected pixels ($\mathrm{DETECT\_MINAREA} = 30$). Objects with $\mathrm{FWHM} \geq 4 \ \mathrm{arcsec}$ were kept. 
Finally, a manual rejection was applied to all images to exclude non-galaxies objects (e.g. a part of a previously masked galaxy, blended compact objects, etc.) and artifacts (e.g., ghosts of optics, edge effects, etc). 
The final catalog contains $6258$ galaxies in the $3^{\circ} \times 3^{\circ}$ Coma field and $3152$ galaxies in the $3^{\circ} \times 3^{\circ}$ control field.

The procedure is illustrated in 
the upper panels of Figure \ref{fig:data}, where we show a $3\arcmin\times 3\arcmin$ section from an image after applying the key steps of the detection pipeline. 
The lower panel show examples of five galaxies from our sample. The galaxies have similar sizes ($r_{\mathrm{eff}} \sim 3.5 \ \mathrm{kpc}$) but span a wide range of luminosities. The faintest one with $m_i=19.82 \ \mathrm{mag}$ is easily detected in the CFHT data. 

\subsection{Structural and Photometric Parameters} 
\label{sec:galfit}
We use SExtractor only to obtain a catalog that includes all large galaxies down to a faint detection limit. Structural and photometric parameters of the galaxies were determined from parametric
fits of the two-dimensional surface brightness distribution, using
GALFIT (\citealt{2002AJ....124..266P}) in two steps. First, fits were performed on the summed $g+i$ images, with neighboring
objects masked,  to determine structural parameters of the galaxies. A single-component S\'ersic profile was assumed, and the values of $r_{\mathrm{eff}}$, $n$, $m_{g+i}$, $\mathrm{b}/\mathrm{a}$, $\mathrm{PA}$ and the sky were fitted by GALFIT. 
The galaxies were then fitted again in each band separately to determine (total) magnitudes and colors. In this second fit all previously free parameters were fixed except the integrated $m_g$ and $m_i$ magnitudes.  A total of 6045 out of 6132 were successfully fit, with average and median $\chi^{2}$ of 1.39 and 1.1, respectively. A small fraction of fits (1.4 \%) did not converge.

\subsection{Red Sequence and Cluster Member Selection} \label{sec:CMD}
The color-magnitude diagrams (CMDs) for galaxies in the Coma field and the control field are shown in Figure \ref{fig:CMD}. Only large galaxies with $r_{\rm eff}>4\arcsec$ are shown, as we are complete above this limit (see above). The red sequence (\citealt{2000AJ....120.2148G})
is very clear in the Coma field, as expected. A visual comparison of the two fields suggests
that most of the bluest and reddest objects in the Coma field are background galaxies, and we
begin by selecting objects close to the red sequence.
We fit the red sequence in the Coma field CMD, using the least squares method and we obtain $(g-i) = - 0.045  M_i + 0.113$. This relation is similar to that obtained by Head et al. (\citeyear{2014MNRAS.440.1690H}).
In the following, we only consider galaxies that are within  $\pm 0.2$ magnitudes of the red sequence line in the analysis. As is evident in the right panel of Figure\ \ref{fig:CMD}, the contamination by unrelated objects is small but non-zero within these limits. 

	\section{Results} \label{sec:results}
\subsection{The Size-Luminosity Plane}

The distribution of red galaxies in the size-magnitude plane, color coded by their number, is shown in Figure \ref{fig:scatter}.
Galaxies are placed into bins of size and magnitude in the $g$ band. The upper left panel shows galaxies in the Coma field, the upper right panel shows galaxies in the control field and the bottom panel shows the subtracted histogram where control galaxies are subtracted from the Coma histogram for each bin of size and magnitude. 
In the upper panels we show galaxies with their apparent sizes and magnitudes (measured in arcseconds and magnitudes, respectively) and in the lower panel we compute their physical sizes and absolute magnitudes under the assumption that they are all at the distance of the Coma cluster ($100 \ \mathrm{Mpc}$).

Galaxies span a wide range of sizes and magnitudes, from dwarf galaxies to UDGs and giant ellipticals. The sample is not complete for dwarf galaxies, due to our cut on $\mathrm{FWHM} \geq 4\arcsec$ which roughly corresponds to an effective radius of $\sim 1 \ \mathrm{kpc}$ at a distance of $100 \ \mathrm{Mpc}$. 
The red symbols denote a population of early-type objects, adapted from Brodie et al.\ (\citeyear{2011AJ....142..199B}). This is not a complete sample, but thought to be representative for different classes of dynamically-hot, early-type objects. 
Galaxies in our Coma sample show a markedly different distribution:
the bright end follows the Brodie et al.\ data points, but there is no fall-off in the number toward fainter magnitudes. 
This conclusion echoes that of van Dokkum et al.\ (\citeyear{2015ApJ...804L..26V}), who showed that UDGs fall in a region of size-luminosity space that has very few previously-known objects.  In this paper, we extend those results and show that there is a continuous magnitude distribution for large objects from the canonical size-luminosity relation all the way to the UDG regime. We note here that our sample contains 18 of the 20
Dragonfly-discovered UDGs that have $\mu_{g,0}<25$\,mag\,arcsec$^{-2}$ in the
van Dokkum et al.\ (2015) sample.\footnote{The two that are missing have $\mu_{g,0}= 24.8$\,mag\,arcsec$^{-2}$ in van Dokkum et al.\ (2015).}


\begin{figure*}[t!]
{\centering
  \includegraphics[width=190mm]{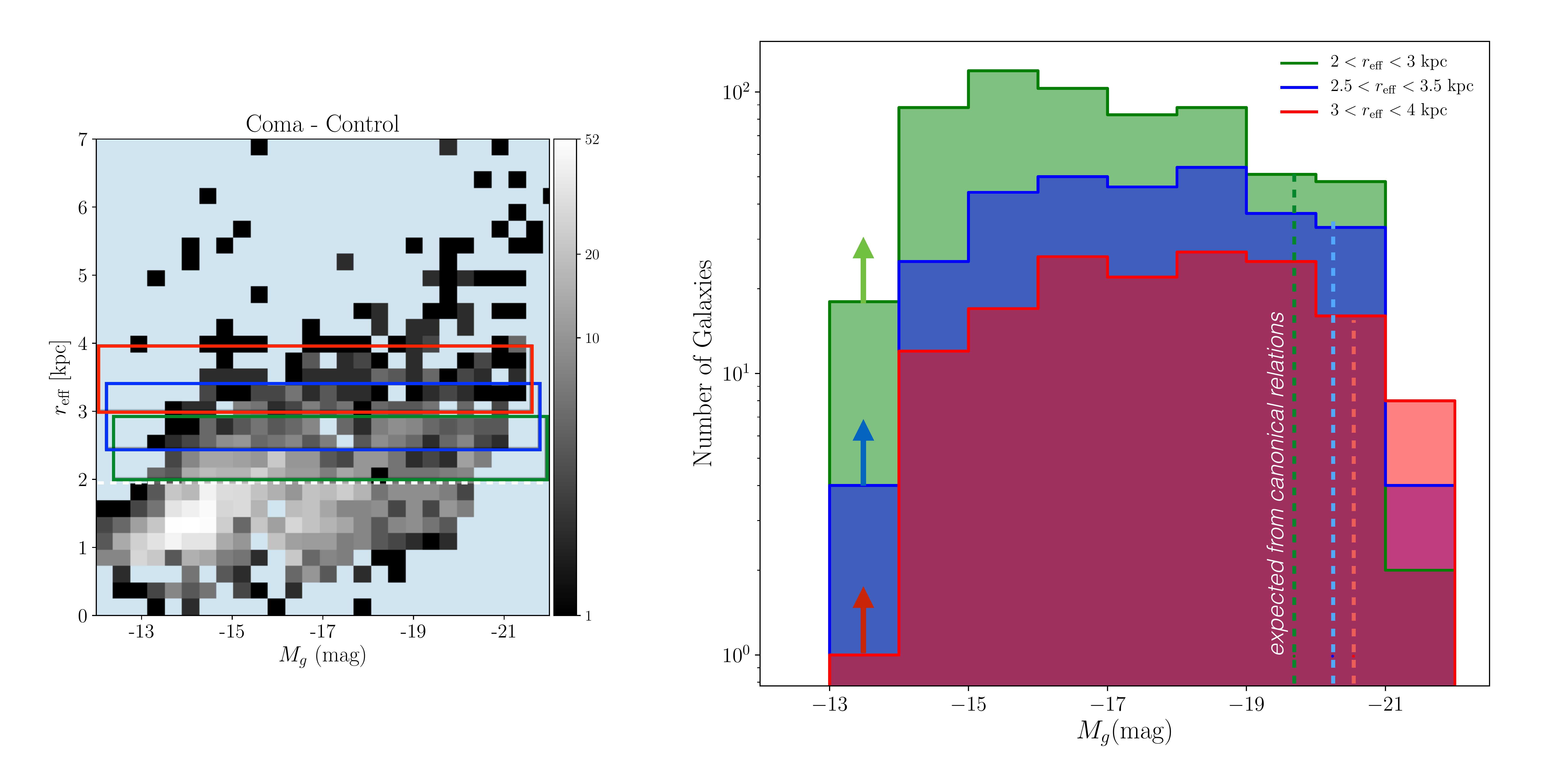}
  \caption{The right panel shows the luminosity function of galaxies in the background subtracted Coma field in three effective radius bins: 2--3\,kpc (green), 2.5--3.5\,kpc (blue) and 3--4\,kpc (red). The bins are indicated in the size-magnitude plane in the left panel. Dashed vertical lines in the right panel mark the mean magnitudes expected from the canonical size-magnitude relation, derived in Bernardi et al.\ (\citeyear{2014MNRAS.443..874B}). Arrows indicate bins that are highly incomplete. The luminosity functions are very broad, with no evidence for a peak at a particular magnitude.}
  \label{fig:hist-cutouts}
  }
\end{figure*}

\subsection{Size-limited sample}

\begin{figure*}[t!]
{\centering
  \includegraphics[width=100mm]{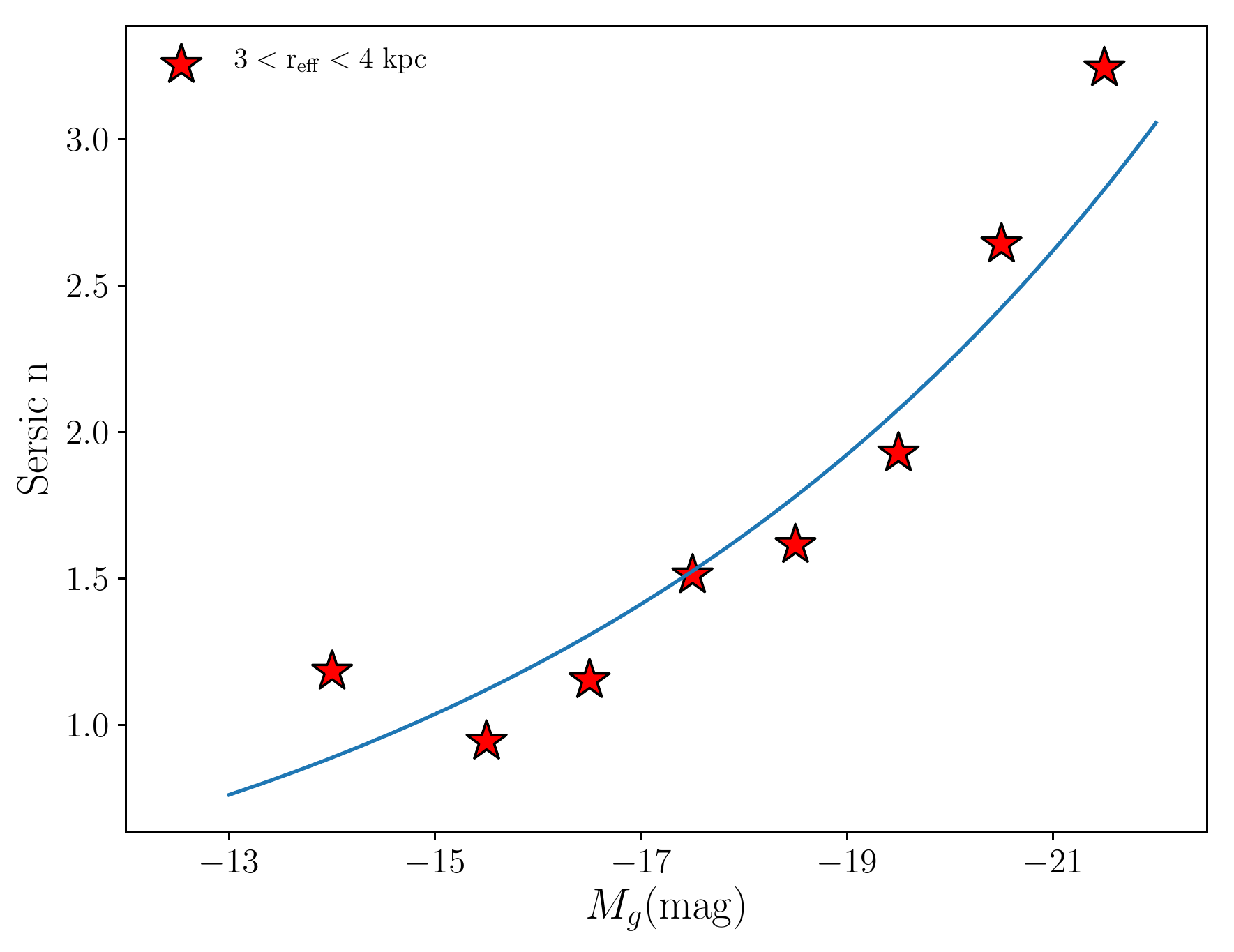}
  \caption{The median S\'ersic index for galaxies in the 3--4\,kpc size bin as a function of magnitude. The line is a power law fit. The structure of large galaxies changes systematically with magnitude. }
  \label{fig:sersic}
  }
\end{figure*}

\citealt{2014MNRAS.443..874B} study the size-luminosity relations for a large sample of $z \leq 0.1$ early-type galaxies by fitting $\sim 5 \times 10^5$ galaxies the from Sloan Digital Sky Survey DR7 Main Galaxy Sample with an apparent magnitude limit of $m_r < 17.75$ mag.
In order to examine how the recently identified large population of UDGs affect the large size end of the size-luminosity relation, we analyze the luminosity distribution of Coma galaxies in three size bins: $r_{\rm eff}=2-3$\,kpc, $r_{\rm eff}=2.5-3.5$\,kpc, and $r_{\rm eff}=3-4$\,kpc. 
The right panel of Figure \ref{fig:hist-cutouts} shows the luminosity function of red cluster galaxies, after subtracting the control field galaxies, in the different size bins.

The distribution of galaxies in all three size bins is broad with sharp cutoffs on both ends. 
The sharp decrease in the number of galaxies for magnitudes fainter than $M_g=-14$ is due to photometric incompleteness, and there is no evidence for a ``preferred" magnitude of large galaxies. The best-fitting logarithmic slopes of the luminosity functions in the three size bins are

\small
\begin{equation} \log_{10}\textrm{N} = \left\{
\begin{array}{@{}rl@{}}
(0.0569 \pm 0.0102)m_g + 0.901 & 2<r_{\textrm{eff}}<3 \ \mathrm{kpc}; \\
(-0.008 \pm 0.0147)m_g + 1.756 & 2.5<r_{\textrm{eff}}<3.5 \ \mathrm{kpc}; \\
(-0.0259 \pm 0.0218)m_g +1.753 & 3<r_{\textrm{eff}}<4 \ \mathrm{kpc}, \\
\end{array}
\right .
\end{equation}
\normalsize

\noindent
in the magnitude range of $-21 < {M}_{g}<-14$. 
A large fraction of the galaxies have a central surface brightness fainter than $24 \ \mathrm{mag} \ \mathrm{arcsec}^{-2}$ in g band which associates them with the UDGs population. 

This result appears to be in tension with previous studies that reported a low measured scatter ($\sigma_{\mathrm{rms}}(\log_{10} r_{\rm eff})<0.15$) around the mean size-luminosity relation (\citealt{2014MNRAS.443..874B}).
Bernardi et al.\ derive a size-luminosity relation for early-type galaxies of the form
\begin{equation}
\log_{10} R = 12.814 + 1.379 M_{r} + 0.038M_{r}^{2} 
\end{equation}
and so for effective radii of 2.5, 3 and $3.5 \ \mathrm{kpc}$, we get absolute magnitudes of $M_{r} = -20.40$, $-20.97$ and $-21.28$ mag, respectively. 
The three dashed lines in the right panel of Figure \ref{fig:hist-cutouts} mark these mean magnitudes for the three size-limited samples, after converting $M_{r}$ to $M_{g}$.
The magnitude distribution does not peak at the expected locations. We stress, however, that the studies are not necessarily in conflict: our study merely shows that the luminosity-size relation (measured here) is very different from the size-luminosity relation (measured previously).

Finally, we show that faint large galaxies are structurally different from bright large galaxies. 
In Figure \ref{fig:sersic} we show the median S\'ersic index $\langle n\rangle$ for galaxies in the $3-4$\,kpc size bin as a function of magnitude. The S\'ersic index monotonically decreases towards fainter magnitudes with median values of $\langle n \rangle \approx 3$ for $M_{g} = -21$ and $\langle n\rangle \sim 1 $ for the low surface brightness galaxies in the sample. The line shows a  fit of the form 
\begin{equation}
\log_{10}\,n = -0.067M_g-0.989.
\end{equation}
This trend is in agreement with the median S\'ersic index of $\langle \mathrm{n} \rangle=0.8$ in Mowla et al.\ (\citeyear{2017ApJ...851...27M}), who determined the structure of UDGs in Coma from deep Subaru images.





\section{Discussion} \label{sec:discuss}
In this Letter, we revisited the luminosity function of galaxies in the Coma cluster using size-limited samples, down to the low surface brightness regime, accounting for the recently discovered population of UDGs. 
We find that the luminosity function for intrinsically-large galaxies is nearly flat across eight orders of magnitudes ($-22 < M_g < -14$), with a slope that is close to  zero for all three samples (Figure \ref{fig:hist-cutouts}). We also find that the S\'ersic index decreases systematically for fainter galaxies, and we infer that the structure of large bright galaxies is very different from large faint galaxies. This is consistent with many other studies, which have shown that large bright galaxies have high S\'ersic indices (\citealt{2009ApJS..182..216K}) and UDGs have morphologies similar to dwarf spheroidals in the Local Group (\citealt{2012AJ....144....4M}, \citealt{2017ApJ...851...27M}, \citealt{2015ApJ...798L..45V}). 
We note that other studies have shown that large faint galaxies are not only different from large bright galaxies but also from small faint galaxies. In particular, the specific frequency of globular clusters in Coma UDGs is a factor of $\sim 7$ higher than that of other galaxies of the same luminosity (\citealt{2016ApJ...819L..20B}, \citealt{2017ApJ...844L..11V}, \citealt{2018ApJ...862...82L}).  

The flat, wide nature of the luminosity function in fixed size bins appears to be in tension with the canonical relation between galaxy size and luminosity across the entire magnitude range. The previously derived tight size-luminosity relation is partly a result of selection effects, as typically only high surface brightness galaxies were considered (e.g., \citealt{2014MNRAS.443..874B}). However, we emphasize that the fraction of faint galaxies that are large is small; that is, UDGs are in the tail of the size distribution at fixed low luminosity. As our data are highly incomplete for galaxies smaller than 2\,kpc, we cannot re-derive the size-luminosity relation. We also note that there may be an environmental dependence, such that the luminosity function of large galaxies could be different in the general field. Van der Burg et al. (\citeyear{2017A&A...607A..79V}) find that the ratio of the number of UDGs to the number of luminous galaxies has a slight dependence on halo mass, such that UDGs are relatively more common in clusters than in groups.

Kravtsov (\citeyear{2013ApJ...764L..31K}) used the abundance matching ansatz to obtain estimates of the virial radius, $R_{200}$, over a wide range of stellar mass and showed that galaxies follow an approximately linear relation between their half-mass radius, $r_{1/2}$ and their virial radius. Since $R_{200}$ scales with the host dark matter halo, $M_{200}$, it implies that the effective radius scales with the halo mass. For $r_{\mathrm{eff}}$ of $\sim 3 \ \mathrm{kpc}$, we obtain a halo mass of $M_{h} \sim 3 \times 10^{12} \ \mathrm{M}_{\odot}$ from the Kravtsov relations. However, we find that at fixed size of $\sim 3 \ \mathrm{kpc}$, galaxies span at least a factor of $\sim 600$ in luminosity. Taken together, these results suggest that at fixed halo mass, there is a factor of $\sim 600$ in stellar mass, which is inconsistent with previous estimates of the scatter in the Stellar Mass-Halo Mass (SMHM) relation (e.g. \citealt{2013ApJ...770...57B}; \citealt{2013MNRAS.428.3121M}) and also with halo occupation statistics (e.g. \citealt{2002ApJ...575..587B}, \citealt{2002MNRAS.329..246B}, \citealt{2018MNRAS.475L.116A}). 
The most straightforward interpretation is that UDGs do not follow  the $r_{1/2} \approx 0.015 \ R_{200}$ relation derived in Kravtsov (\citeyear{2013ApJ...764L..31K}). An interesting implication is that the baryons in UDGs ``fill'' a larger fraction of the volume of their halos than other galaxies of the same luminosity; this likely explains their high $M/L$ ratios within their half-light radii (e.g. \citealt{2016ApJ...819L..20B}, \citealt{2016ApJ...828L...6V}, \citealt{2018ApJ...856L..31T}). 


In conclusion, UDGs are an important population in galaxy clusters, and are far more common than large galaxies on the canonical size-luminosity relation. Specifically, the number of galaxies with $2\,$kpc\,$<r_{\rm eff}<3$\,kpc that fall
within $\pm 0.5$ magnitudes of the size-luminosity relation is a factor of 9 smaller than those with $-19<M_g<-13$. Deeper imaging is needed to determine whether the luminosity function remains flat at even fainter magnitudes. Furthermore, extending the study to smaller galaxies would enable a complete study of the size-luminosity plane. Finally, we confirm that UDGs are not a distinct population in the size-luminosity plane. Previous studies have shown that there is no bimodality in the sizes of galaxies at fixed luminosity (e.g. \citealt{2018RNAAS...2a..43C}); here we make the complementary point that there is no bimodality in luminosity at fixed size.

\acknowledgments

The authors thank Charlie Conroy for valuable discussions.
Support from NSF grants AST-1312376 and AST-1613582 is gratefully acknowledged.

\bibliography{coma_ref} 

\end{document}